\documentclass[prd,aps,nofootinbib,onecolumn,showpacs,10pt]{revtex4-1}
\usepackage{graphicx,epsfig}

\usepackage{epsf}
\usepackage{graphicx}

\begin{document}

\title{  \bf    Investigation of heavy-heavy pseudoscalar mesons in  thermal QCD Sum Rules}
\author{ E. Veli Veliev $^{*1}$,
K. Azizi $^{\dag2}$, H. Sundu $^{*3}$, N. Ak\c sit$^{\ddagger4}$ \\
$^{*}$Department of Physics , Kocaeli University, 41380 Izmit,
Turkey\\
 $^{\dag}$Physics Division,  Faculty of Arts and Sciences,
Do\u gu\c s University,
 Ac{\i}badem-Kad{\i}k\"oy, \\ 34722 Istanbul, Turkey\\
 $^{\ddagger}$Faculty of Education , Kocaeli University, 41380 Izmit,
Turkey\\
 $^1$ e-mail:elsen@kocaeli.edu.tr\\
$^2$e-mail:kazizi@dogus.edu.tr\\
$^3$email:hayriye.sundu@kocaeli.edu.tr\\
$^4$email:nurcanaksit@kocaeli.edu.tr}

\begin{abstract}
We investigate the mass and decay constant of the  heavy-heavy pseudoscalar, $B_c$,
$\eta_c$ and  $\eta_b$  mesons in the framework of
 finite temperature  QCD sum rules. The
annihilation and scattering parts of spectral density are calculated
in the lowest order of perturbation theory. Taking into account the
additional operators arising at finite temperature, the
nonperturbative corrections are also evaluated.  The masses and
decay constants remain unchanged under $T\cong 100 ~MeV$, but after
this point, they start to diminish with increasing the temperature.
At  critical or deconfinement temperature, the decay constants reach
approximately to 38\% of their values in the vacuum, while the
masses are decreased about 5\%, 10\% and 2\%  for $B_c$, $\eta_c$
and $\eta_b$ states, respectively. The results at zero  temperature
are in a good consistency with the existing experimental values as
well as predictions of the other nonperturbative approaches.
\end{abstract}

\maketitle

\section{Introduction}
Over the last two decades, there is an increasing interest on
properties of hadrons under extreme conditions \cite{K.Yagi,
J.Letessier}. According to these investigations, two theoretical
aspects, namely  theoretical studies of hadrons
at finite temperature and density as well as a careful analysis of the
heavy ion collision results  are important.  Calculation of hadronic parameters at
finite temperature and density directly from  QCD is a
difficult problem. The thermal QCD is successful theory in the large
momentum transfer region, where the quark-gluon running coupling
constant is small and one can reliably use perturbative approaches.
However, at the hadronic scale, this coupling constant becomes large and perturbation theories fail. Hence,  investigation of hadronic properties requires some
nonperturbative approaches. Some nonperturbative approaches
are  lattice QCD, heavy quark effective theory (HQET), different quark models and QCD sum
rules.  Among these approaches, the QCD sum rule method
\cite{M.A.Shifman} and its extension to the finite temperature
\cite{A.I.Bochkarev} has been extensively used as an efficient tool
to hadron physics \cite{P.Colangelo}. The same as QCD sum rules in vacuum, the main idea in thermal QCD
sum rules also is to relate the hadronic parameters  with the
QCD degrees of freedom. In this method, an appreciate thermal correlator is expressed in
terms of interpolating currents of participating particles. From one side, this correlation function is evaluated saturating it by a tower of hadrons with the same
 quantum numbers as the interpolating currents. On the other hand,
it is calculated via the operator product
expansion (OPE) in terms of operators having different mass dimensions. Matching these two different representations of the same correlation function
provides us a possibility to predict hadronic properties in terms of
finite-temperature perturbation theory and long-distance
nonperturbative physics including the thermal quark and
gluon condensates as well as thermal average of energy density.

Comparing to the QCD sum rules in vacuum, the thermal QCD sum rules have several new features. One of them is to take into account  the
interaction of the currents with the existing particles in the medium. Such interactions require modification of the  hadronic spectral function. The other
aspect is  breakdown of Lorentz invariance by the choice of
reference frame. Due to residual $O(3)$ symmetry at finite temperature,  more operators with
the same dimensions appear in the OPE  compared
to those at zero temperature \cite{E.V.Shuryak,T.Hatsuda,S.Mallik}.
The thermal QCD sum rule method has been extensively used to
study the thermal properties of light
\cite{S.Mallik1,S.Mallik2,E.V.Veliev}, heavy-light
\cite{C.A.Dominguez1,C.A.Dominguez2,E.V.Veliev1} and heavy-heavy
\cite{F.Klingl,K.Morita,K.Morita1,E.V.Veliev2} mesons as a reliable
and well-established method.

The discussion of heavy mesons properties at zero temperature has a
rather long history
\cite{V.A.Novikov,T.M.Aliev,L.J.Reinders,I.I.Balitsky,C.A.Dominguez,T.M.Aliev1,
S.S.Gershtein,P.Ball,S.Narison,V.V.Kiselev,M.Jamin,TWChiu,W.Wang,T.M.Aliev2,T.M.Aliev3}.
The heavy mesons play very important role in our understanding of
nonperturbative dynamics of QCD. First determinations of leptonic
decay constant of pseudoscalar, $B_c$ meson at zero temperature were
made twenty years ago \cite{T.M.Aliev1,S.S.Gershtein}. Such charged
meson decays play important role to extract the magnitudes and
phases of the Cabbibo-Kobayashi-Maskawa(CKM) matrix elements, which
can help us understand the origins of CP violation in and beyond the
standard model.
 Our aim in
this work is to investigate the temperature dependence of mass and
leptonic decay constants of the pseudoscalar  $B_c$, $\eta_c$ and $\eta_b$ mesons taking
into account the additional operators arising at finite temperature. The pseudoscalar decay constant, $f_P$ is defined by vacuum to meson matrix
element of the axial vector current as:

\begin{eqnarray}\label{nolabel}
{\langle}0|(\overline{Q}_1 \gamma_{\mu} \gamma_5 Q_2)(0)|P
{\rangle}=if_P q_{\mu},
\end{eqnarray}
where $Q_{1,2}=c$ or $b$ and $P=B_c, \eta_c$ or $ \eta_b$. In thermal field
theories, the  meson mass, $m_P$ and its decay constant, $f_P$
should be replaced by their temperature dependent versions.

The paper is organized as follows. In section 2, we obtain thermal QCD  sum rules for the masses and decay constants of the considered pseudoscalar mesons calculating the
spectral densities and  nonperturbative corrections. In section 3,  we present our numerical calculations and  discussions.

\section{Thermal QCD Sum Rules for Decay Constants and Masses of Heavy Pseudoscalar,  $B_c, \eta_c$  and $ \eta_b$ Mesons}

Taking into account the new aspects of the  finite temperature QCD,
 sum rules for the masses and  decay constants of the heavy
pseudoscalar mesons containing $b$ and/or $c$ quark are derived in this section.  The starting point is to consider the following responsible   two-point
thermal correlation function:
\begin{eqnarray}\label{correl.func.1}
\Pi(q,T) =i\int d^{4}xe^{iq \cdot x}{\langle} {\cal T}\left ( J^{P}
(x)  J^{P\dag}(0)\right){\rangle},
\end{eqnarray}
where $T$ denotes the temperature,  ${\cal T}$  is the time ordering
product and $J^{P}(x)=\overline{Q}_{1}(x)i\gamma_{5}Q_{2}(x)$ is the
interpolating current of the heavy pseudoscalar mesons. The thermal
average of the operator, $A= {\cal T}\left ( J^{P}
(x)  J^{P\dag}(0)\right)$ appearing in the above correlation function is expressed as:
\begin{eqnarray}\label{A}
{\langle}A {\rangle}=\frac{Tr(e^{-\beta H} A)}{Tr( e^{-\beta H})},
\end{eqnarray}
where $H$ is the QCD Hamiltonian,  $\beta = 1/T$ is  inverse
of the temperature $T$ and traces are performed over any complete
set of states.

As we previously mentioned, to obtain sum rules for physical observables, we need to calculate the aforementioned correlation function in two different ways. In QCD or theoretical side, the correlation function is calculated in deep
Euclidean region, $q^2\ll-\Lambda_{QCD}^2$ via OPE where the short
or perturbative and long distance or nonperturbative contributions are
separated,
\begin{eqnarray}\label{correl.func.QCD1}
\Pi^{QCD}(q,T) =\Pi^{pert}(q,T)+\Pi^{nonpert}(q,T).
\end{eqnarray}
The perturbative contribution   is  calculated using
perturbation theory, whereas the nonperturbative   contributions are
expressed in terms of the thermal expectation values of the quark
and gluon condensates as well as  thermal average of the energy density. The
perturbative part can be written in terms of a dispersion integral, hence
\begin{eqnarray}\label{correl.func.QCD1}
\Pi^{QCD}(q,T) =\int \frac{ds \rho(s,T)}{s-q^2}+\Pi^{nonpert}(q,T),
\end{eqnarray}
where, $\rho(s,T)$  is called the spectral density at finite
temperature.  The thermal spectral density at fixed $|\bf{q}|$ is written as:
\begin{eqnarray}\label{rhoq}
\rho(q,T)=\frac{1}{\pi}~Im\Pi^{pert}(q,T)~\tanh\left(\frac{\beta
q_{0}}{2}\right).
\end{eqnarray}

In order to calculate the $\rho(q,T)$ in the lowest
order in perturbation theory, we use quark propagator at finite
temperature \cite{A.Das} as:
\begin{eqnarray}\label{prop}
S(q)=(\gamma_{\mu}~q^{\mu}+m)\left(\frac{1}{q^2-m^2+i\varepsilon}+2\pi i
n(|q_{0}|)~\delta(q^2-m^2)\right),
\end{eqnarray}
where $n(x)=\left[ exp
(\beta x)+1 \right ]^{-1}$ is the Fermi distribution function. Using the above propagator, after some
calculations we find  the imaginary part of the correlation
function as:
\begin{eqnarray}\label{eqn7}
Im\Pi(q,T)=L(q_0)+L(-q_0),
\end{eqnarray}
where,
\begin{eqnarray}\label{eqn8}
L(q_0)&=&-N_c\int\frac{d
\textbf{k}}{8\pi^2}\frac{\omega_1^2-\textbf{k}^2+\textbf{k}\cdot \textbf{q}-
\omega_1
q_0-m_1m_2}{\omega_1\omega_2}\Big\{\Big[\Big(1-n_1(\omega_1)\Big)\Big(1-n_2(\omega_2)\Big)+n_1(\omega_1)n_2(\omega_2)\Big]
\delta(q_0-\omega_1-\omega_2)
\nonumber\\
&-&
\Big[\Big(1-n_1(\omega_1)\Big)n_2(\omega_2)+\Big(1-n_2(\omega_2)\Big)n_1(\omega_1)\Big]\delta(q_0-\omega_1+\omega_2)\Big\}.
\end{eqnarray}
Here, $\omega_1=\sqrt{\textbf{k}^2+m_1^2}$ and
$\omega_2=\sqrt{\textbf{(q-k)}^2+m_2^2}$. As it is seen, the
$L(q_0)$ involves two pieces. The first
term, which includes delta function $\delta(q_0-\omega_1-\omega_2)$
survives at zero temperature and is called the annihilation term.
The second term, which includes delta function
$\delta(q_0-\omega_1+\omega_2)$ is called scattering term and
vanishes at $T=0$. The delta function,  $\delta(q_0-\omega_1-\omega_2)$
in Eq. (\ref{eqn8}) gives the first branch cut, $q^2\geq(m_1+m_2)^2$,
which coincides with zero temperature cut that describes the
standard threshold for particle decays. On the other hand, the delta
function, $\delta(q_0-\omega_1+\omega_2)$ in Eq. (\ref{eqn8}) shows
that an additional branch cut arise at finite temperature,
$q^2\leq(m_1-m_2)^2$, which corresponds to particle absorption from
the medium. Taking into account these contributions, the annihilation
and scattering parts of spectral density in the case, $\textbf{q}=0$
can be written as:

\begin{eqnarray}\label{Rhoa}
\rho^{a,pert}(s,T)=\rho_{0}(s)
\Big[1-n\Big(\frac{\sqrt{s}}{2}\Big(1+\frac{m_1^2-m_2^2}{s}\Big)\Big)-
n\Big(\frac{\sqrt{s}}{2}\Big(1-\frac{m_1^2-m_2^2}{s}\Big)\Big)
\Big],
\end{eqnarray}
for $(m_1+m_2)^2\leq s\leq \infty$, and
\begin{eqnarray}\label{Rhos}
\rho^{s,pert}(s,T)=\rho_{0}(s)
\Big[n\Big(\frac{\sqrt{s}}{2}\Big(1+\frac{m_1^2-m_2^2}{s}\Big)\Big)-
n\Big(-\frac{\sqrt{s}}{2}\Big(1-\frac{m_1^2-m_2^2}{s}\Big)\Big)
\Big],
\end{eqnarray}
for $0\leq s\leq (m_1-m_2)^2$ with $m_1\geq m_2$. Here $\rho_0(s)$, is the
spectral density in the lowest order of perturbation theory at zero
temperature and  is given by:
\begin{eqnarray}\label{Rho0s}
\rho_{0}(s)=\frac{3}{8\pi^2s}q^2(s)v(s),
\end{eqnarray}
where $q(s)=s-(m_1-m_2)^2$ and $v(s)=\sqrt{1-4m_1m_2/q(s)}$.

In our calculations, we also take into account the  perturbative
two-loop order $\alpha_{s}$ correction to the spectral density. For
equal quark masses case this correction at zero temperature can be
written as \cite{L.J.Reinders}:
\begin{eqnarray}\label{Pi1}
\rho_{\alpha_{s}}(s)= \frac{s \alpha_{s} v }{2\pi^3}\Big[
\frac{\pi^2}{2v}-\frac{1+v}{2}\Big(\frac{\pi^2}{2}-3\Big)+F(v)\ln\frac{1+v}{1-v}+G(v)\Big],
\end{eqnarray}
where  $F(v)$ and $G(v)$ functions have the following forms:
\begin{eqnarray}\label{Pi11}
F(v)=\frac{3}{4v^3}-\frac{21}{16v}-\frac{18v}{16}+\frac{3v^3}{16},
\end{eqnarray}
and
\begin{eqnarray}\label{Pi12}
G(v)=-\frac{3}{2v^2}+\frac{9}{8}-\frac{3v^2}{8}.
\end{eqnarray}
Here $v=v(s)$ and  we replace the strong coupling $\alpha_{s}$ in
Eq. (\ref{Pi1}) with its temperature dependent lattice improved
expression \cite{K.Morita,O.Kaczmarek}. When doing the numerical
calculations for $B_c$ meson, the contribution coming from two-loop
diagrams is used for unequal quark masses case $\rho_{\alpha_{s}}$
\cite{L.J.Reinders, C.A.Dominguez}, but since its expression is very
lengthy, we do not present its explicit expression here.

To calculate the  nonperturbative part in QCD side, we use the nonperturbative part of the quark propagator in an
external gluon field, $A^a_{\mu}(x)$ in the Fock-Schwinger gauge,
$x^{\mu}A^a_{\mu}(x)=0 $. Taking into account one  and two
gluon lines attached to the quark line, the massive quark propagator
in momentum space can be written as \cite{L.J.Reinders}:

\begin{eqnarray}\label{Saap}
S^{aa^{\prime}}(k)&=&\frac{i}{\not\!k-m}\delta^{aa^{\prime}}
-\frac{i}{4}g (t^{c})^{aa^{\prime}}
G^{c}_{\kappa\lambda}(0)\frac{1}{(k^2-m^2)^2}\Big[\sigma_{\kappa\lambda}
(\not\!k+m)+(\not\!k+m)\sigma_{\kappa\lambda}\Big]
\nonumber\\
&-&\frac{i}{4} g^2
(t^{c}t^{d})^{aa^{\prime}}G^{c}_{\alpha\beta}(0)G^{d}_{\mu\nu}(0)
\frac{\not\!k+m}{(k^2-m^2)^5}(f_{\alpha\beta\mu\nu}+f_{\alpha\mu\beta\nu}
+f_{\alpha\mu\nu\beta})(\not\!k+m),
\end{eqnarray}
where,
\begin{eqnarray}\label{f}
f_{\alpha\beta\mu\nu}=\gamma_{\alpha}(\not\!k+m)\gamma_{\beta}(\not\!k+m)
\gamma_{\mu}(\not\!k+m)\gamma_{\nu}.
\end{eqnarray}
In order to proceed, we also  need to know the expectation value,
$\langle Tr G_{\alpha\beta}G_{\mu\nu}\rangle$. The Lorentz
covariance at finite temperature allows us to write the general
structure of this expectation value in the following way:
\begin{eqnarray}\label{TrGG} \langle Tr^c G_{\alpha \beta} G_{\mu \nu}
\rangle &=& \frac{1}{24} (g_{\alpha \mu} g_{\beta \nu} -g_{\alpha
\nu} g_{\beta \mu})\langle G^a_{\lambda \sigma} G^{a \lambda \sigma}\rangle \nonumber \\
 &+&\frac{1}{6}\Big[g_{\alpha \mu}
g_{\beta \nu} -g_{\alpha \nu} g_{\beta \mu}-2(u_{\alpha} u_{\mu}
g_{\beta \nu} -u_{\alpha} u_{\nu} g_{\beta \mu} -u_{\beta} u_{\mu}
g_{\alpha \nu} +u_{\beta} u_{\nu} g_{\alpha \mu})\Big]\langle
u^{\lambda} {\Theta}^g _{\lambda \sigma} u^{\sigma}\rangle,
\end{eqnarray}
where,   $u^{\mu}$  is  the four-velocity of the heat bath and it is introduced to restore Lorentz invariance formally in the thermal field theory.
In the rest  frame of the heat bath,  $ u^{\mu} = (1, 0, 0, 0)$ and $u^2 = 1$. Also $\Theta^{g}_{\lambda \sigma}$ is   the traceless, gluonic part of the stress-tensor of the QCD.
 Therefore, up to terms necessary for our calculations, the non perturbative part of  massive
quark propagator at finite temperature takes the form:
\begin{eqnarray}\label{Saap1}
S^{aa^{\prime }nonpert}(k)&=&
 -\frac{i}{4}g (t^{c})^{aa^{\prime}}
\frac{G^{c}_{\kappa\lambda}}{(k^2-m^2)^2}
\Big[\sigma_{\kappa\lambda}
(\not\!k+m)+(\not\!k+m)\sigma_{\kappa\lambda}\Big]
+\frac{i~g^2~\delta^{aa^{\prime}}} {9~(k^2-m^2)^4}\Big\{\frac{3m
(k^2+m \not\!k)}{4}\langle
G^{c}_{\alpha\beta}G^{c\alpha\beta}\rangle
\nonumber\\
& +& \Big[ m \Big(k^2-4(k\cdot u)^2\Big) +\Big(m^2 -4(k\cdot
u)^2\Big)\not\!k +4(k\cdot u)(k^2-m^2)\not\!u\Big]\langle
u^{\alpha}\Theta^{g}_{\alpha\beta}u^{\beta}\rangle\Big\},\end{eqnarray}
Using the above expression and after straightforward
calculations, the nonperturbative part  in QCD side is  obtained as:
\begin{eqnarray}\label{NonPert}
\Pi^{nonpert}&=& \int_{0}^{1}~dx\Big\{
\frac{\langle\alpha_{s}G^{2}\rangle}{48 \pi \Big[m_1^2
(-1+x)-\Big(m_2^2+q^2 (-1+x)\Big)x\Big]^4}\Big[-m_1^6 (-1+x)^4
(5-20x+3x^2)+m_1^5 m_2 (-1+x)^2
\nonumber\\
&\times& (5-6x+12x^2-14x^3+6x^4)-x^4 \Big(24~m_2^2~ q^4
(-1+x)^3+6~q^6 ~(-1+x)^4+m_2^6~(-12+14x+3x^2)
\nonumber\\
&+&2~m_2^4~ q^2~(15-31x+16x^2)\Big)+m_1^4 (-1+x)^2 x
\Big(-2~q^2(-1+x)^2(-1+16x)+m_2^2 (-8+58x-56x^2+3x^3) \Big)
\nonumber\\
&+& m_1^2 (-1+x)x^2 \Big(24~q^4~(-1+x)^3 x +m_2^4
(3-45x+47x^2+3x^3)+3~m_2^2~q^2~(-1+23x-44x^2+22x^3)\Big)
\nonumber\\
&-&m_1^3 m_2 (-1+x)x
\Big(q^2~(-8+5x+9x^2-7x^3+x^4)+2~m_2^2~(4-3x+9x^2-12x^3+6x^4)\Big)
+m_1~m_2~x^2
\nonumber\\
&\times& \Big(3~q^4(1-5x^2+6x^3-2x^4)-m_2^2~q^2~(6-9x^2+2x^3+x^4)
+m_2^4~(3+6x^2-10x^3+6x^4)\Big) \Big]
\nonumber\\
&+&\frac{\alpha_{s}\langle
u^{\alpha}\Theta^{g}_{\alpha\beta}u^{\beta}\rangle}{72 \pi
\Big[m_1^2 (-1+x)-\Big(m_2^2+q^2 (-1+x)\Big)x\Big]^4}\Big[6~m_1^6
(-1+u^2)(-1+x)^4(5-20x+3x^2)-6~m_1^5~m_2(-1+u^2)
\nonumber\\
&\times& (-1+x)^2
(5-6x+12x^2-14x^3+6x^4)+2~m_1^2~(-1+x)x^2\Big[-3~m_2^4~(-1+u^2)
(3-45x+47x^2+3x^3)
\nonumber\\
&-& m_2^2~q^2~(-1+x)\Big(-9(1-22x+22x^2)+u^2(23-227x+227x^2)
\Big)+q^4(-1+x)^2\Big(72(-1+x)x+u^2(-14+96x
\nonumber\\
&-& 92x^2+x^3)\Big) \Big]-m_1^4 (-1+x)^2 x \Big(6~m_2^2 (-1+u^2)
(-8+58x-56x^2+3x^3)+q^2(-1+x)(12~(1-17x+16x^2)
\nonumber\\
&+& u^2(-37+259x-239x^2+8x^3))
\Big)+x^3\Big[6~m_2^6~(-1+u^2)~x~(-12+14x+3x^2)+q^6 (-1+x)^3
\Big(-36(-1+x)x
\nonumber\\
&+& u^2(9-47x+47x^2)\Big)+2~m_2^2~q^4~(-1+x)^2
\Big(-72(-1+x)x+u^2(9-85x+89x^2+x^3)\Big)+m_2^4~q^2~(-1+x)
\nonumber\\
&\times& \Big(12(15-16x)x+u^2(9-195x+215x^2+8x^3)\Big)
\Big]+2~m_1^3~m_2~(-1+x)x\Big[6~m_2^2~(-1+u^2)(4-3x+9x^2-12x^3
\nonumber\\
&+& 6x^4)
+q^2(-1+x)\Big(-3(8+3x-6x^2+x^3)+2~u^2(12+x-8x^3+4x^4)\Big)
\Big]-2~m_1~m_2~x^2
\Big[3~m_2^4(-1+u^2)
\nonumber\\
&\times& (3+6x^2-10x^3+6x^4)+q^2(-1+x)^2\Big(9(-1-2x+2x^2)
+u^2(9+14x-12x^2-4x^3+2x^4)\Big)+m_2^2~q^2~(-1+x)
\nonumber\\
&\times& \Big(2(-6-6x+3x^2+x^3) +2~u^2
(9+7x-8x^3+4x^4)\Big)\Big]-4(-1+x)x\Big\{2~m_1^3~m_2~(-1+x)^2~x~
(7-11x+8x^2)\nonumber\\
&-& m_1^4(-1+x)^2
(-25+55x-47x^2+8x^3)-2~m_1~m_2~(-1+x)x^2\Big(m_2^2(4-5x+8x^2)+2q^2
(-2+3x-2x^2+x^3)\Big)
\nonumber\\
&+& 2~m_1^2~(-1+x)~x~\Big(m_2^2(-14+29x-29x^2)
+q^2(14-38x+44x^2-21x^3+x^4)\Big)+x^2\Big(q^4(-1+x)^2(9-11x
\nonumber\\
&+& 11x^2)
+m_2^4(9-15x+23x^2+8x^3)+2~m_2^2~q^2(-9+22x-30x^2+16x^3+x^4)\Big)
\Big\}~(q\cdot u)^2 \Big] \Big\},
\end{eqnarray}
where, $\langle
G^{2}\rangle=\langle
G^{c}_{\alpha\beta}G^{c\alpha\beta}\rangle$.

Now, we turn our attention to the physical or phenomenological side of the correlation function. The hadronic spectral density is expressed by the ground state
pseudoscalar meson pole plus the contribution of the  higher states and continuum. According to quark-hadron duality, the continuum is expected to
be well approximated by the QCD spectral density calculated in
perturbation theory starting at some threshold $s_0$. Therefore, the
hadronic spectral density can be written as:
\begin{eqnarray}\label{RhoHads}
\rho^{had}(s)=\frac{f_P^2(T)m_P^4(T)}{(m_1+m_2)^2}\delta(s-m_P^2)+
\theta(s-s_0)\rho^{pert}(s)
\end{eqnarray}

Matching the phenomenological and QCD sides of the correlation
function,  sum rules for the mass and decay constant of
pseudoscalar meson are obtained. To suppress the contribution of the
higher states and continuum, the Borel transformation over the $q^2$
as well as continuum subtraction are performed. As a result of the
above procedure and after lengthy calculations, we obtain the following sum rule for the
decay constant:
\begin{eqnarray}\label{fPs}
f_P^2(T)~
m_P^4(T)~e^{-\frac{m_P^2}{M^2}}=(m_1+m_2)^2~\Big\{\int_{(m_1+m_2)^2}^{s_0(T)}
ds~\rho^{a,pert}(s)~e^{-\frac{s}{M^2}}+\int_{0}^{(m_1-m_2)^2}
ds~\rho^{s,pert}(s)~e^{-\frac{s}{M^2}}+\widehat{B}\Pi^{nonpert}\Big\},\nonumber\\
\end{eqnarray}
where $M^2$ is the Borel mass parameter.

The sum rule for the mass is obtained applying derivative with
respect to $-\frac{1}{M^2}$ to the both sides of the sum rule for
the decay constant of the pseudoscalar meson in Eq. (\ref{fPs}) and dividing by itself:
\begin{eqnarray}\label{mPs}
m_P^2(T)=\frac{\int_{(m_1+m_2)^2}^{s_0(T)}
ds~\rho^{a,pert}(s)~s~\exp(-\frac{s}{M^2})+\int_{0}^{(m_1-m_2)^2}
ds~\rho^{s,pert}(s)~s~\exp(-\frac{s}{M^2})+{\Pi_1}^{nonpert}(M^2,T)}{\int_{(m_1+m_2)^2}^{s_0(T)}
ds~\rho^{a,pert}(s)~\exp(-\frac{s}{M^2})+\int_{0}^{(m_1-m_2)^2}
ds~\rho^{s,pert}(s)~\exp(-\frac{s}{M^2})+\widehat{B}\Pi^{nonpert}},
\end{eqnarray}
where,
\begin{eqnarray}\label{Pinp}
{\Pi_1}^{nonpert}(M^2,T)=M^4\frac{d}{dM^2}\widehat{B}\Pi^{nonpert},
\end{eqnarray}
 and $\widehat{B}\Pi^{nonpert}$ shows the nonperturbative part of QCD side in Borel transformed scheme and is given
by:

\begin{eqnarray}\label{BorelNonPert}
\hat{B}\Pi^{nonpert}&=&\int_{0}^{1}dx~\frac{1}
{96~\pi~M^6~x^4~(-1+x)^4}
\exp\Big[\frac{m_2^2x-m_1^2(-1+x)}{M^2x(-1+x)}\Big]
\Big\{\langle\alpha_{s}G^{2} \rangle \Big[-m_1^6 (-1+x)^6+m_1^5 m_2
(-1+x)^4~x
\nonumber\\
&\times&(-1+2x)+ x^4\Big(-12~m_2^2~M^4~(-1+x)^3+12~M^6~(-1+x)^4
+2~m_2^4~M^2~x~(-1+x)-m_2^6 x^2\Big)
\nonumber\\
&+&m_1^4 x(-1+x)^3 \Big(2 M^2(-1+x)^2+m_2^2 (1-3x+x^2)\Big)+m_1^2
x^2 (-1+x)\Big(12 M^4 x (-1+x)^3+m_2^4 x (-1+x+x^2)
\nonumber\\
&+& 3m_2^2 M^2(1-3x+4x^2-2x^3)\Big)+m_1^3 m_2 x (-1+x)^2 \Big(-m_2^2
x(1-2x)^2+M^2(2-9x+6x^2+x^3)\Big)\nonumber\\
&-&m_1 m_2(-1+x)
 x^2
\Big(m_2^4x^2(1-2x)-m_2^2 M^2 x (6-9x+x^2)+6 M^4
(1+x-4x^2+2x^3)\Big) \Big]
\nonumber\\
&+& 3~\alpha_{s}\langle
\Theta^{g}\rangle
\Big[m_1^6(-1+x)^6-m_1^5 m_2 x (-1+x)^4 (-1+2x)+m_1~m_2 x^3
(-1+x)\Big(m_2^4 x(1-2x)
\nonumber\\
&+&4~M^4~
 (-1+x)^2(2-x+x^2)+m_2^2~M^2~
 (-4+3x+5x^2-4x^3)\Big)-m_1^4 x
(-1+x)^3 \Big(m_2^2 (1-3x+x^2)
\nonumber\\
&+&2M^2
 (1-2x+x^3)\Big)-m_1^2 x^2 (-1+x)
 \Big(m_2^4 x
(-1+x+x^2)+m_2^2 M^2 (5-17x+24x^2-12x^3)
\nonumber\\
&+&
 M^4 (-1+x)^2
 (-1+15x-7x^2+2x^3)\Big)+x^3 \Big(m_2^6 x^3 +M^6
(-1+x)^3(9-11x+11x^2)+2m_2^4 M^2 x
\nonumber\\
&\times&  (-1+4x-4x^2+x^3)-m_2^2 M^4 (-1+x)^2 (-9+7x
+x^2+2x^3)\Big)+m_1^3 m_2 x^2 (-1+x)^2
\nonumber\\
 &\times&\Big(m_2^2 (1-2x)^2+M^2
(1+6x-11x^2+4x^3)\Big) \Big] \Big\}, \nonumber \\
\end{eqnarray}

 where, $\Theta^{g}=\Theta^{g}_{00}$. Following \cite{E.V.Veliev2},  we also use the gluonic part of energy density both obtained from lattice QCD
 \cite{M.Cheng,D.E.Miller} and chiral perturbation theory \cite{P.Gerber}. In the rest
  frame of the heat bath, the results of some observables calculated using
 lattice QCD in  \cite{M.Cheng}
 are fitted well by the following  parametrization for the thermal average of total energy density, $\langle \Theta \rangle$:
\begin{eqnarray}\label{tetag}
\langle \Theta \rangle= 2 \langle \Theta^{g}\rangle=
6\times10^{-6}exp[80(T-0.1)](GeV^4),
\end{eqnarray}
where temperature $T $ is measured in units of $GeV$ and this
parametrization is valid only in the interval   $0.1~GeV\leq T \leq
0.17~GeV$. Here, we would like to stress that the total energy density has been calculated for
$T\geq 0$ in chiral perturbation theory, while this quantity has
only been obtained for $T\geq 100~MeV$  in lattice QCD (for details see
\cite{M.Cheng,D.E.Miller}). In low temperature chiral perturbation
limit, the thermal average of the  energy density is expressed as  \cite{P.Gerber}:
\begin{eqnarray}\label{tetagchiral}
\langle \Theta\rangle= \langle \Theta^{\mu}_{\mu}\rangle +3~p,
\end{eqnarray}
where, $\langle
\Theta^{\mu}_{\mu}\rangle$ is trace of the total energy momentum
tensor and $p$ is pressure. These quantities are given by:
\begin{eqnarray}\label{tetamumu}
\langle
\Theta^{\mu}_{\mu}\rangle=\frac{\pi^2}{270}\frac{T^{8}}{F_{\pi}^{4}}
\ln \Big(\frac{\Lambda_{p}}{T}\Big), ~~~~~~~~~~~~~~p=
3~T~\Big(\frac{m_{\pi}~T}{2~\pi}\Big)^{\frac{3}{2}}\Big(1+\frac{15~T}{8~m_{\pi}}+\frac{105~T^{2}}{128~
m_{\pi}^{2}}\Big)exp\Big(-\frac{m_{\pi}}{T}\Big),
\end{eqnarray}
where $\Lambda_{p}=0.275~GeV$, $F_{\pi}=0.093~GeV$ and
$m_{\pi}=0.14~GeV$.

In our calculation we use the temperature dependent continuum
threshold, $s_0(T)$, gluon condensate, $\langle G^2\rangle$ and
 strong coupling constant,  as presented in \cite{E.V.Veliev2} (for details see \cite{C.A.Dominguez2,M.Cheng,D.E.Miller,K.Morita,O.Kaczmarek}).

\section{Numerical analysis}

In this section, we numerically  analysis
the sum rules for the  masses and decay constants of the heavy-heavy
pseudoscalar mesons. We use the values,
$m_c=(1.3\pm0.05)GeV$, $m_b=(4.7\pm0.1)GeV$ and ${\langle}0\mid
\frac{1}{\pi}\alpha_s G^2 \mid 0 {\rangle}=(0.012\pm0.004)GeV^4$ for quark masses and gluon condensate at zero temperature.
From the  sum rules  for the masses and decay constants it is  clear that they also contain two
auxiliary parameters, namely  continuum threshold. $s_0$ and Borel mass
parameter, $M^2$ as the main inputs. These are not physical quantities, hence the physical observables should be independent of these parameters. Therefore, we should look for working regions for these parameters
 at which the dependence of the masses and decay constants on these parameters is weak.  The continuum threshold, $s_{0}$ is not
completely arbitrary,  but it is  in correlation with the energy of
the first exited state with the same quantum numbers as the
considered interpolating currents.  We choose the values  $ 44~
GeV^2 \leq s_0 \leq 46~ GeV^2 $,  $ 11~ GeV^2 \leq s_0 \leq 12~
GeV^2 $ and  $ 94~ GeV^2 \leq s_0 \leq 97~ GeV^2 $  for the
continuum threshold in accordance with  $B_c$, $\eta_c$ and
$\eta_{b}$ channels, respectively. The working region for the Borel
mass parameter, $M^2$ is determined as following. Its lower limit is
calculated requiring that the higher states and continuum
contributions constitute approximately 30\%
 of the total dispersion integral. Its upper limit is obtained demanding that
the mass sum rules  should be convergent, i.e., contribution
of the operators with higher dimensions is small. As a result of the above
procedure, the working region for the Borel parameter is found to be
$ 10~ GeV^2 \leq M^2 \leq 25~ GeV^2 $,  $ 6~ GeV^2 \leq M^2 \leq 12~
GeV^2 $ and $ 15~ GeV^2 \leq M^2 \leq 30~ GeV^2 $ in  $B_c$,
$\eta_c$ and $\eta_{b}$ channels, respectively.

\begin{figure}[h!]
\begin{center}
\includegraphics[width=12cm]{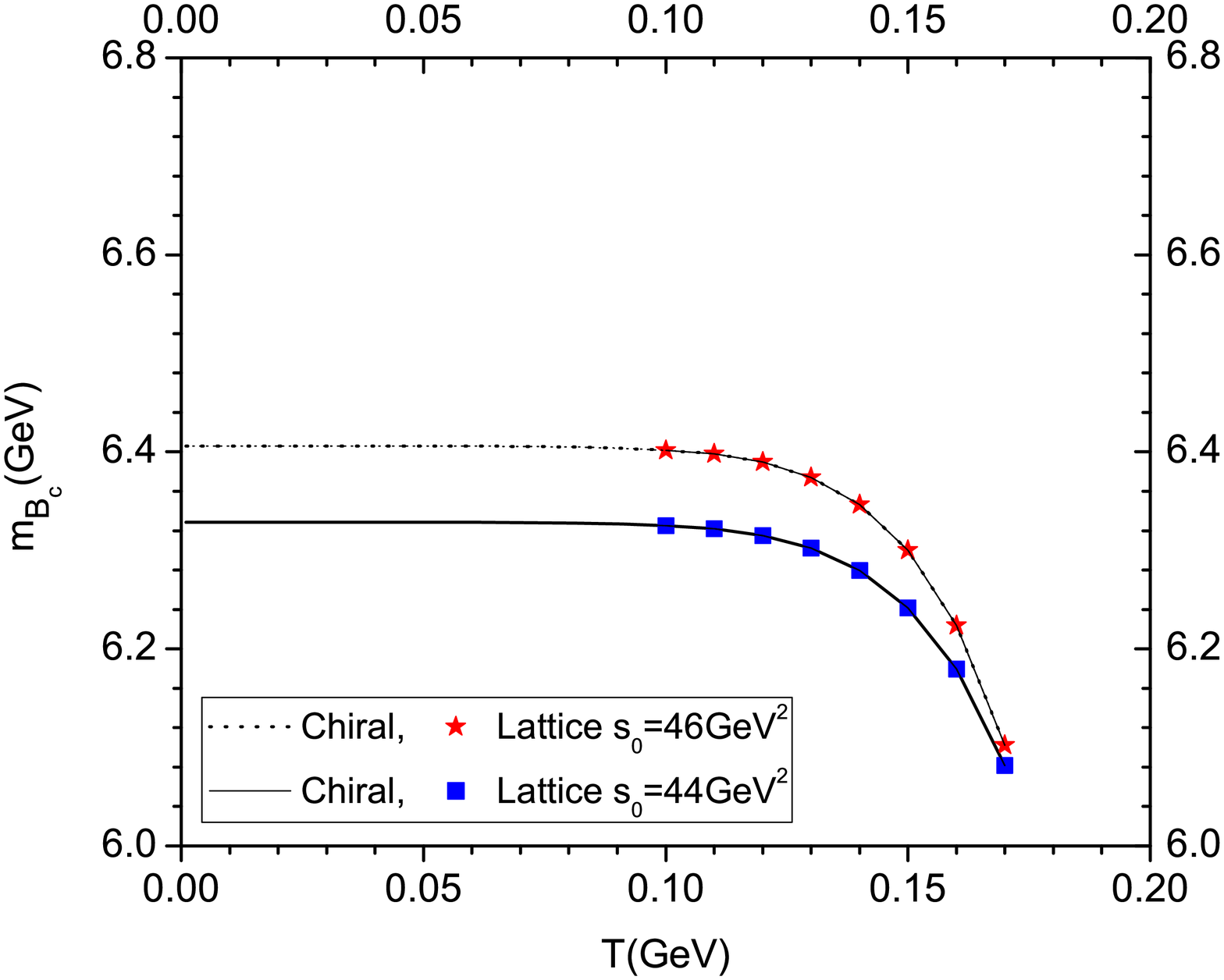}
\end{center}
\caption{The dependence of the mass of $B_c$  meson  on temperature
for Chiral and Lattice QCD parametrization of the gluonic part of
the energy density.} \label{mBcTemp}
\end{figure}

\begin{figure}[h!]
\begin{center}
\includegraphics[width=12cm]{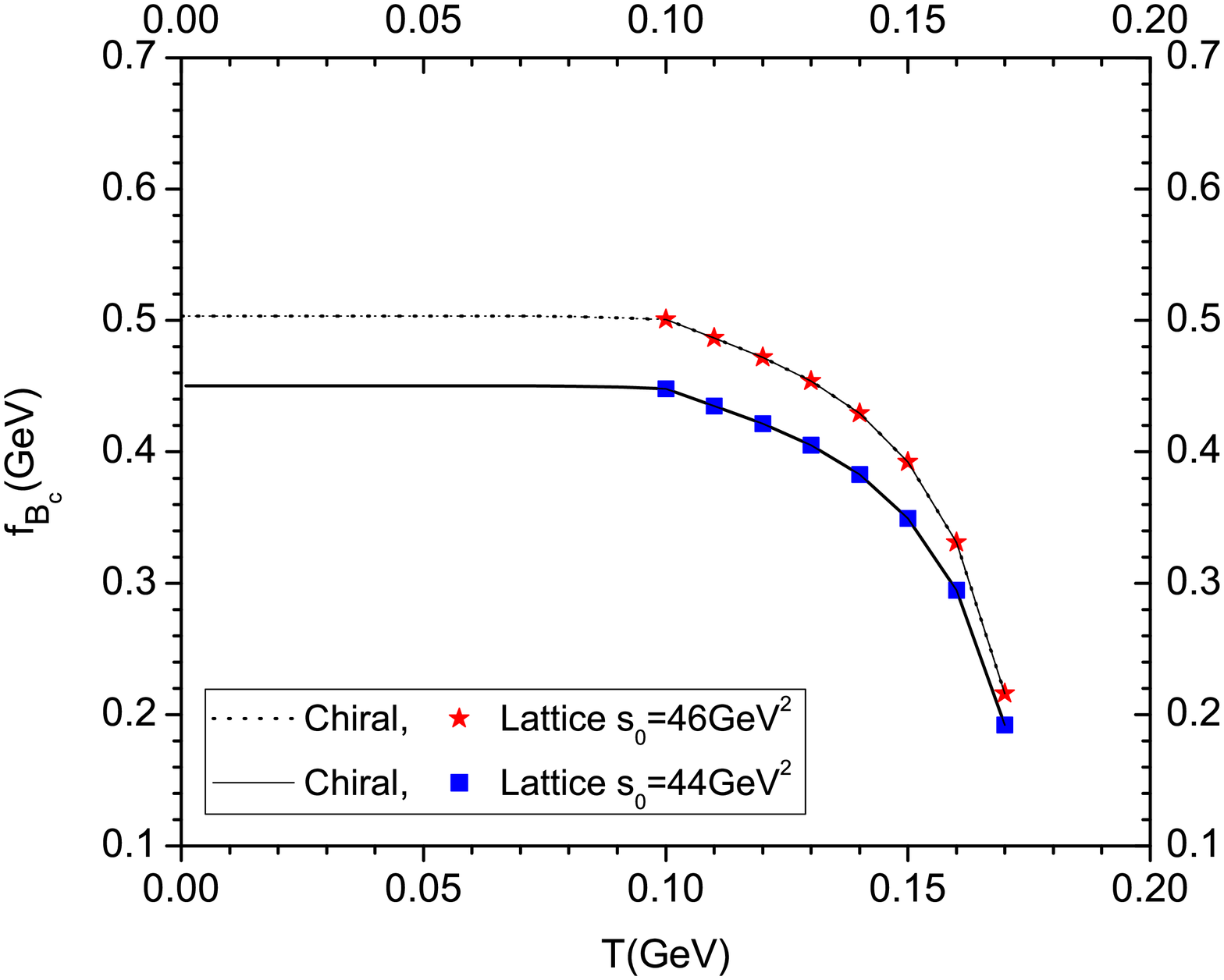}
\end{center}
\caption{The same as Fig.  \ref{mBcTemp} but for $f_{B_c}$.}  \label{fBcTemp}
\end{figure}

\begin{figure}[h!]
\begin{center}
\includegraphics[width=12cm]{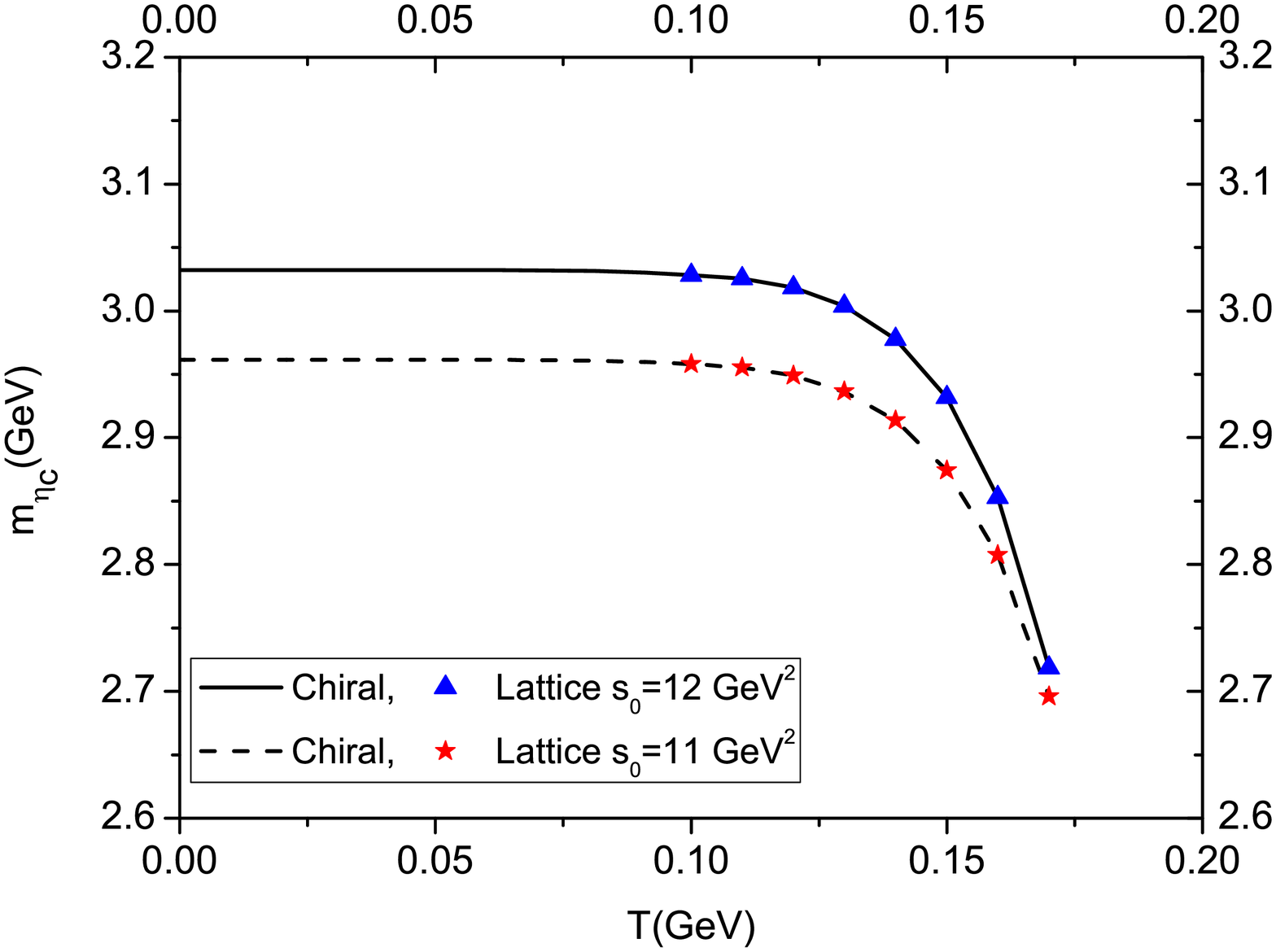}
\end{center}
\caption{The same as Fig.  \ref{mBcTemp} but for $m_{\eta_c}$.}
\label{metacTemp}
\end{figure}

\begin{figure}[h!]
\begin{center}
\includegraphics[width=12cm]{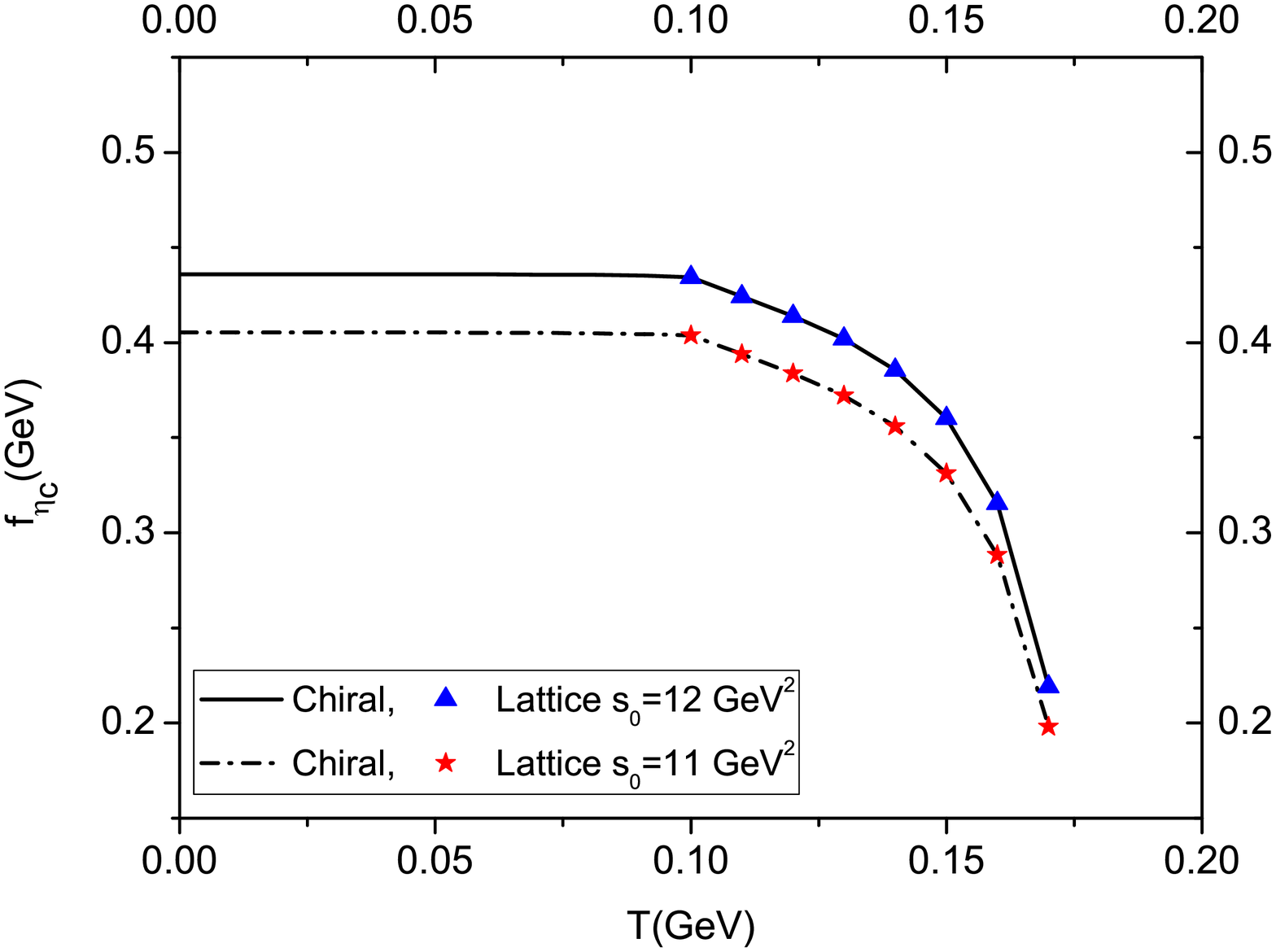}
\end{center}
\caption{The same as Fig.  \ref{mBcTemp} but for $f_{\eta_c}$.}
\label{fetacTemp}
\end{figure}

\begin{figure}[h!]
\begin{center}
\includegraphics[width=12cm]{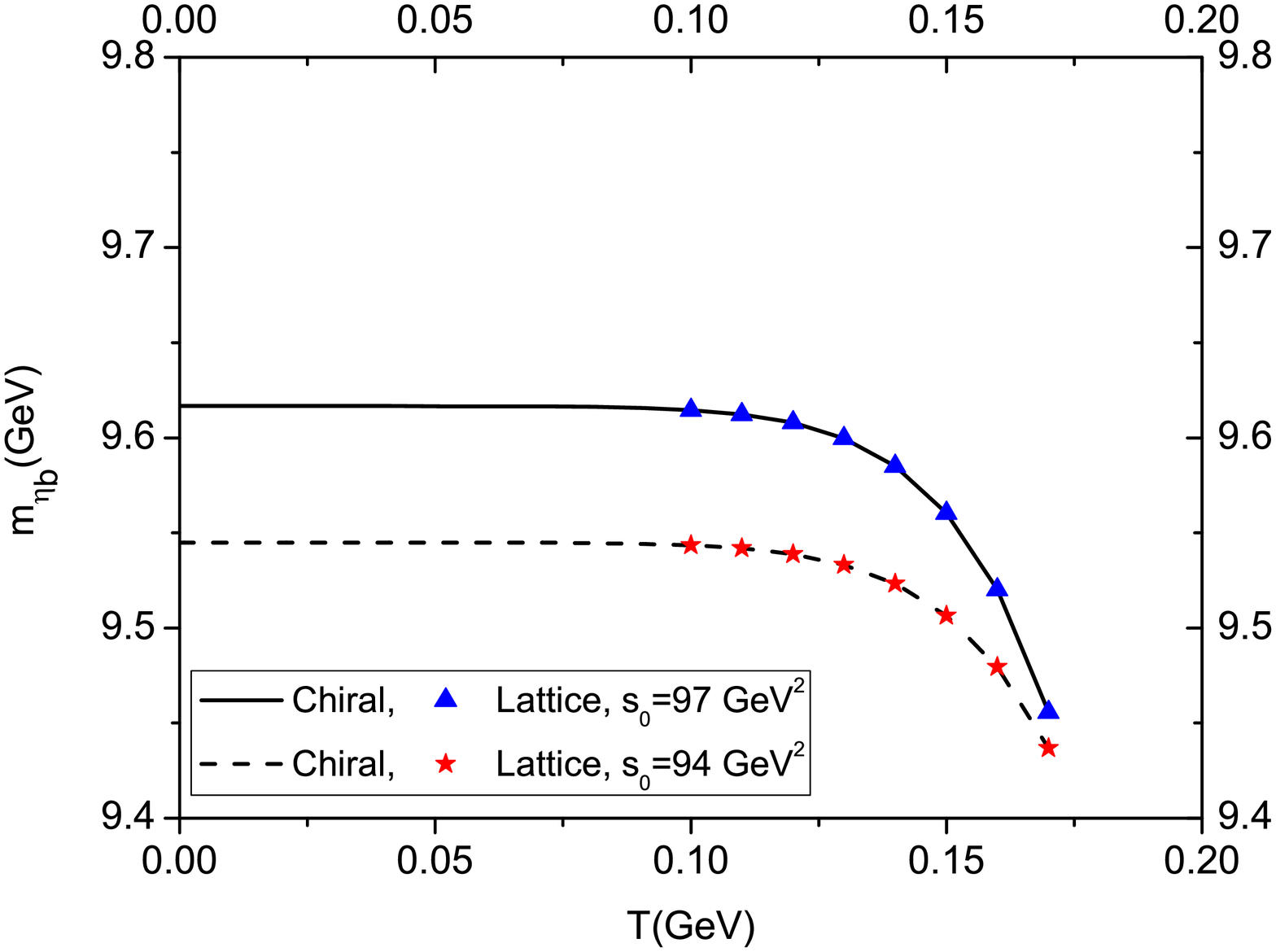}
\end{center}
\caption{The same as Fig.  \ref{mBcTemp} but for $m_{\eta_b}$.}
\label{metabTemp}
\end{figure}

\begin{figure}[h!]
\begin{center}
\includegraphics[width=12cm]{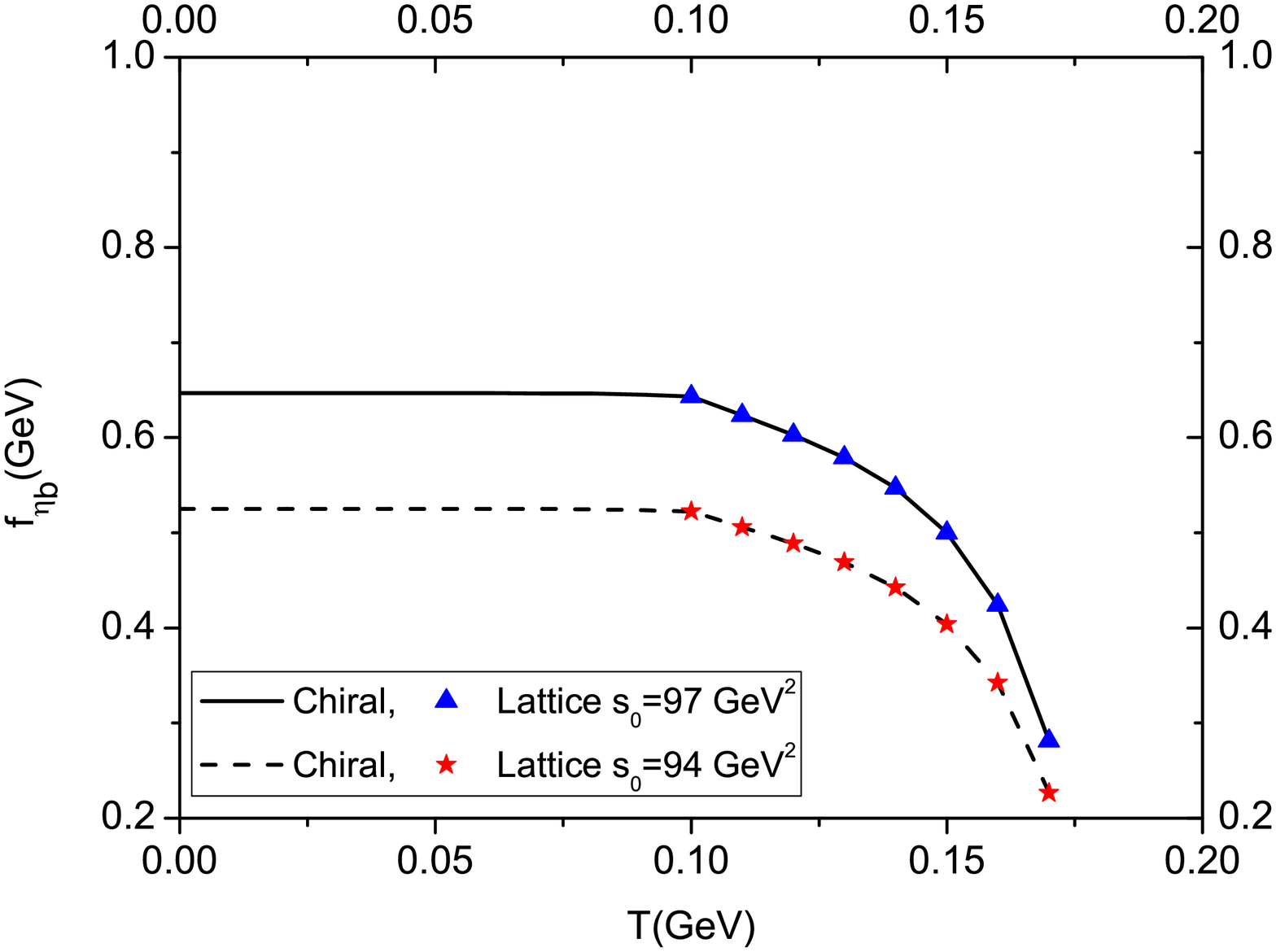}
\end{center}
\caption{The same as Fig.  \ref{mBcTemp} but for $f_{\eta_b}$.}
\label{fetabTemp}
\end{figure}

Our calculations show that in the working regions the dependence of
the considered observables on auxiliary parameters is weak. We 
depict the dependence of masses and decay constants
 on the temperature, $T$ in Figs.
\ref{mBcTemp}-\ref{fetabTemp}.  These figures contain the results
obtained using both lattice QCD and chiral perturbation parametrization
for  the gluonic part of the energy density. These figures depict that
both parametrization of lattice QCD and chiral perturbation  theory predict the same result in validation  limit of lattice QCD fit parametrization, i.e., $0.10~GeV\leq T \leq
0.17~GeV$. These figures also show
 that
the masses and decay constants remain unchanged approximately up to $T\simeq100~MeV$,
 but after this point, they start to diminish with increasing the
temperature. Near the critical or deconfinement
temperature,
 the decay constants reach approximately to 38\% of their values in vacuum,
  while the masses are decreased about 5\%, 10\%, 2\%  comparing with their values at zero temperature for $B_c$,
  $\eta_c$,
$\eta_{b}$ mesons, respectively.
 From these figures, we obtain the results on the  decay constants and masses  at zero temperature as presented
 in Tables I and II. The quoted errors in these Tables are due to the errors in variation of the
 continuum threshold at zero temperature, Borel mass parameter as well as errors coming from fit parametrization of the temperature dependent continuum threshold, gluon condensate and strong coupling constant
and uncertainties existing in other input parameters. These Tables
also include  the existing predictions of the other works as well as
experimental data. The Table II depicts a very good consistency
between our results and the experimental data on masses but from
Table I, we see that the present work results and the results
existing in the literature (see Table I) on the decay constant are
comparable up to presented errors.

  Our results for
the leptonic decay constants at zero temperature  as well as the
behavior of the masses and decay constants of the considered
pseudoscalar heavy mesons  with respect to the temperature can be
checked in the future experiments. The obtained behavior of the
observables in terms of temperature  can be used in analysis of the
results of the heavy ion collision experiments.

\begin{table}[h]
\renewcommand{\arraystretch}{1.5}
\addtolength{\arraycolsep}{3pt}
$$
\begin{array}{|c|c|c|c|}
\hline \hline
         &f_{B_c}(MeV) & f_{\eta_{c}}(MeV) &  f_{\eta_{b}}(MeV)   \\
\hline
  \mbox{Present Work}        &  476\pm27   &  421\pm35& 586\pm61 \\
\hline
  \mbox{QCD sum rules \cite{V.A.Novikov,T.M.Aliev1,S.S.Gershtein}}        & 400\pm25  &  350&- \\
\hline
\mbox{Potential Model \cite{V.V.Kiselev}}        & 400\pm45  &  402&599 \\
\hline
\mbox{Lattice QCD Method \cite{TWChiu}}    & 489\pm7  &  438\pm11& 801\pm12 \\
\hline
  \mbox{Experiment \cite{K.W.Edwards}}        & -  &  335\pm75&-\\
                    \hline \hline
\end{array}
$$
\caption{Values of the leptonic decay constants of the heavy-heavy
pseudoscalar, $B_c$, $\eta_{c}$ and $\eta_{b}$ mesons in vacuum.
These results have been obtained using the values
$M^{2}=15~GeV^{2}$, $M^{2}=6~GeV^{2}$ and $M^{2}=20~GeV^{2}$for
$B_c$, $\eta_c$ and
   $\eta_b$ particles, respectively. }
\label{tab:lepdecconst}
\renewcommand{\arraystretch}{1}
\addtolength{\arraycolsep}{-1.0pt}
\end{table}
\begin{table}[h]
\renewcommand{\arraystretch}{1.5}
\addtolength{\arraycolsep}{3pt}
$$
\begin{array}{|c|c|c|c|}
\hline \hline
 &  m_{B_c}~(GeV)& m_{\eta_{c}}~(GeV)& m_{\eta_{b}}~(GeV)\\
\hline
  \mbox{Present Work }       &  6.37\pm0.05   &  2.99\pm0.04& 9.58\pm0.03 \\
\hline
 \mbox{Experiment \cite{C.Amsler}} &  6.277\pm 0.006 &2.9803\pm0.0012 & 9.3909\pm 0.0028  \\
 \hline \hline
\end{array}
$$
\caption{Values of the mass of the heavy-heavy pseudoscalar, $B_c$,
$\eta_{c}$ and $\eta_{b}$ mesons in vacuum. The same values as Table I  for the auxiliary parameters have been used. } \label{tab:mass}
\renewcommand{\arraystretch}{1}
\addtolength{\arraycolsep}{-1.0pt}
\end{table}

\section{Acknowledgement}

The authors would like to thank T. M. Aliev for his useful
discussions. This work is supported in part by the scientific and
technological research council of turkey (TUBITAK) under the
research project no. 110T284 and  research fund of kocaeli
university under grant no. 2011/029.


\end{document}